\documentclass{an}
\usepackage{graphicx}
\usepackage{times}
\usepackage{fancyhdr}
\def\jrasc{JRASC}
\def\aap{A\&A}
\begin{document}

\title{The  Initial Mass Function
 as given by the fragmentation}

\author{Lorenzo Zaninetti}
\institute{Dipartimento di Fisica Generale, Via Pietro Giuria 1,\\
           10125 Torino, Italy}

\date{Received; accepted; published online}

\abstract{The dichotomy  between    a  universal mass function (IMF)
and   a variable IMF which  depends  on   local physical
parameters  characterises
observational and theoretical stellar astronomy.
In this contribution the available distributions of probability
are briefly reviewed. The physical nature of two of them,
gamma variate and lognormal, is then explained once
the framework of the fragmentation is introduced.
Interpolating techniques are then applied to the sample of
the first 10 pc and to  the open cluster NGC6649:
in both cases  lognormal distribution  produces the best
fit.
The three power law function has also been investigated
and visual  comparison  with an artificially generated
sample of 100000 stars suggests that the variations
in the spectral index are simply due to the small number of stars
available in the observational sample.
In order to derive the sample of masses,
 a new formula 
that allows us to express the mass as a function of the 
absolute magnitude and (B-V) for  
MAIN  V  ,  GIANTS III and  SUPERGIANTS I is derived. 
}

\correspondence{zaninetti@ph.unito.it}

\maketitle

\section{Introduction}
The first analytical expression adopted to explain the low mass
distribution of stars, see~\cite{Salpeter}, was a power
law of the type
$\xi ( {\mathcal{M}}) \propto   {\mathcal{M}}^{-\alpha}$
where $\xi   ( {\mathcal{M}})$ represents  the probability
of having  a mass between $ {\mathcal{M}}$ and
$ {\mathcal{M}}+d{\mathcal{M}}$ ; this is called initial mass
  function, in the following IMF\@.
    In the range
$10 \mathcal {M}_{\sun}~>~\mathcal {M} \geq  1 \mathcal {M}_{\sun}$,
 the value of $\alpha$, 2.35, has changed  little
over the decades and a recent evaluation quotes 2.3, see~\cite{Kroupa_2001}.
A fractional exponential law
aimed at explaining  the mass distribution of asteroids,
stars and galaxies
was then introduced by~\cite{Kiang_stars}~.
A three power law  function was later introduced
by~\cite{Scalo}, \cite{Kroupa_1993}, and \cite{Binney} ;
in this case the range of existence
of the  segments as well as the exponent $\alpha$
are functions of the   investigated environment.
 After 15 years,  the number  of segments was raised to four
 once a  universal  IMF was introduced,
 see~\cite{Kroupa_2001}.
 Recent developments translate the concept of
 IMF from our surroundings ( for example the first 10 pc)
 to open clusters such as the Pleiades,
 see~\cite{Kroupa_2004}.
 A connection between   three segment  zones
 and three  physical processes has been investigated
 in \cite{Elmegreen}; these three zones correspond  to
 brown dwarf masses $\approx~0.02\mathcal {M}_{\sun}$,
 to intermediate mass stars and high mass stars.
 The question of uniformity or variability of the initial
 mass function has been carefully explored,
 see~\cite{Meusinger}, \cite{Kroupa_2002},\cite{Shadmehri},
 and \cite{Elmegreen}.

  This paper  briefly  reviews the 
  physical nature of gamma-variate and lognormal  distribution
  (Section~\ref{sec_frag}),
  introduces an algorithm that  deduces the mass from
  the two basic photometric parameters B-V and $M_{\mathrm V}$
  (Section~\ref{sec_algo}),
  and analyses the IMF in the light of  physical
  and non physical functions (Section~\ref{sec_masses}).

\section{Preliminaries}

\label{sec_prelimaries}
 Once  the histogram of the star's  masses is obtained, 
 we can fit it  with three  
 different probability density functions ( in the following pdf):
 two of them are well known , the  lognormal and the power law ,
 while the third is  a gamma-variate  which has been
 applied here for the first time.
 The lognormal distribution    $f_{\mathrm{LN}}(x;\mu,\sigma$)  
 is   characterised   by
the  average value ,$\mu$,
 and the  variance , $\sigma^2$, see~\cite{Evans}.
 The power law distribution   has the form 
 $p(x;\alpha) \propto x^{-\alpha}$
 where $\alpha$ $>$ 0.
A refinement of the power law function 
consists in the multiple--part power law
\begin{equation}
\xi (x) \propto x^{-\alpha_i}
\quad , 
\end{equation}
with i varying from 3 to 4.
This operation means that the field of existence of the
function is broken into  i independents zones.

The gamma-variate is analysed in Section~\ref{sec_frag}.

\subsection {Fragmentation}
\label{sec_frag}

The stellar mass has been viewed as the Jeans mass
in the cloud core, see for example
\cite{Larson},
\cite{Elmegreen_1997},
\cite{Elmegreen_1999},
\cite{Chabrier}, and
\cite{Bate}.
Here conversely we adopt the point of view
of the  fragmentation that was started by~\cite{Kiang}
where a rigorous demonstration of the size distribution
of random Voronoi segments was given.
The Kiang pdf ( a gamma-variate)   has the form 
\begin{equation}
H (x ;c ) = \frac {c} {\Gamma (c)} (cx )^{c-1} e^{-cx}
\quad , \label{kiang}
\end{equation}
where $  0 \leq x < \infty $, $ c~>0$,  and  $\Gamma (c)$ is the
gamma function with argument c
, see~\cite{Kiang}. 
This pdf is characterised by
average value, $\mu$=1, and variance ,$\sigma^2$=1/c.

Starting from a  rigorous result in 1D,  \cite{Kiang} conjectured
that the Voronoi cells in 2D/3D have an area/volume distribution
represented by  equation~(\ref{kiang}) with c=4/6.

We now briefly review a useful  result that can be obtained
once  $x_1 \cdots x_N$ ,independent random variables,
 are introduced.
Supposing that 
\begin{equation}
y = x_1 \times x_2 \times \cdots \times x_{N-1} \times x_{N} =
\prod_{j=1}^N x_j
\quad,
\end{equation}
it can be shown that for large N the pdf of y is approximately
lognormal, see~\cite{Hwei}.
From this point of view the area with $y=x_1\times x_2$
or the volumes $y=x_1 \times x_2 \times x_3$ can be fitted with a lognormal
once the number 2 and 3 are considered a transition to large N.

\section{Mass determination from colour absolute-magnitude}

\label{masses}
\label{sec_algo}
 The  (B-V) colour can be expressed as
\begin{equation}
(B-V)=  K_{\mathrm{BV}}   + \frac {T_{\mathrm{BV}}}{T}
\quad,
\label{bvt}
\end {equation}
here T is  the temperature, $K_{\mathrm{BV}}$  and $T_{\mathrm{BV}}$
are two parameters that  can be derived by implementing
the least square method  on a series of calibrated   data.
 The bolometric correction, BC, can be expressed   as
\begin {equation}
BC  = M_{\mathrm{bol}}  -M_{\mathrm{V}} =
-\frac{T_{\mathrm{BC}}}{T} - 10~\log_{10}~T + K_{\mathrm {BC}}
\quad,
\label{bct}
\end {equation}
where $M_{\mathrm{bol}}$   is the absolute bolometric magnitude,
      $M_{V}$     is the absolute visual     magnitude,
      $T_{\mathrm {BC}}$ and  $K_{\mathrm{BC}}$ are two parameters that can be derived
      through the general linear least square  method
      applied to a series of calibrated data.
We now justify the theoretical basis of  equations~(\ref{bvt})
 and  (\ref{bct}). We can start
from the definition of (B-V)
\begin{equation}
(B-V)=m_{\mathrm B} - m_{\mathrm V}  = K  - 2.5  \log_{10}
\frac
{\int S_{\mathrm B} I_{\lambda} d\lambda}
{\int S_{\mathrm V} I_{\lambda} d\lambda}
\quad,
\label{bv}
\end {equation}
where  $S_{\lambda}$ is the sensitivity function in the region
specified by the index $\lambda$~,
K is a constant 
  and  $I_{\lambda}$ is the energy flux
reaching the earth.
We now define a sensitivity function for a pseudo-monochromatic
BV system
\begin {equation}
S_{\lambda} = \delta (\lambda -\lambda_i)
\quad  i=B,V
\quad,
\end {equation}
where $\delta$ denotes the Dirac delta function.
On inserting the energy flux as given by the Planck
distribution,
the following is found
\begin  {equation}
(B-V)= K\prime
 - {\frac {hc}{kT} ( \frac {\lambda_{\mathrm B} - \lambda_{\mathrm V}}
{\lambda_{\mathrm B} \lambda_{\mathrm V}})}
2.5 \log_{10}( {\mathrm { e}}  )
\label{planck}
\quad,
\end {equation}
where $K\prime$ is a constant ,
c is the velocity of light    , 
k is the Boltzmann's constant ,
h is the Planck's constant    ,
$\lambda_{\mathrm B}$ the wavelength of the filter B,
$\lambda_{\mathrm V}$ the wavelength of the filter V, 
   and T the temperature.
On inserting  $\lambda_{\mathrm B}$ = 4450 \AA~and $\lambda_{\mathrm V}$ = 5500 \AA~in 
equation~(\ref{planck}),
the equation~(\ref{bvt})
is theoretically derived.

A theoretical couple, $T_{\mathrm{BV}}$=6701\,K  and $K_{\mathrm{BV}}$=-0.51
can be found by inserting the constants
of the Planck distribution
in equation~(\ref{planck}) and calibrating the relationship
on the Sun couple, $T_{\sun}$ =5777\,K and $(B-V)_{\sun}$=0.65~.
In a similar way the  bolometric correction is found to be
\begin{equation}
BC =
M_{\mathrm {bol}}  -M_{\mathrm V}=
K^{\prime\prime}  - 2.5  \log_{10}
\frac
{\int _0 ^{\infty}  I_{\lambda} d\lambda}
{\int S_{\mathrm V} I_{\lambda} d\lambda}
\quad,
\label{bc_teo}
\end {equation}
where $K^{\prime\prime}$ is another constant.

When the numerical value of the
exponential in the Planck function is much greater
than one,  equation~(\ref{bct})
is found with the theoretical value $T_{\mathrm {BC}}$=28402\,K
and $K_{\mathrm {BC}}$=42.45~, once  BC of the sun is requested,
$BC_{\sun}$ =-0.08 at $T_{\sun}=5777$\,K.

The application of  equation~(\ref{bvt}) and  (\ref{bct})
to the calibrated physical parameters for stars
of the various   luminosity  classes has been reported
in Table~\ref{coefficients}.

\begin{figure}
\resizebox{\hsize}{!}
{
\includegraphics[scale=0.1,angle=0]{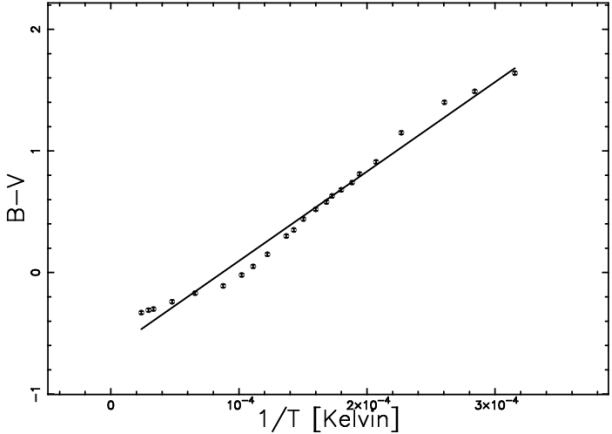}
}
\caption{Least square fit of (B-V) against 1/T, MAIN SEQUENCE V.
         Error on (B-V)=0.025.
         The calibrated data, extracted
         from    Table 15.7 in~\cite{Cox}, are represented with their
         error, and
         the full line reports the suggested fit.}
\label{f01}
\end{figure}

\begin{figure}
\resizebox{\hsize}{!}
{
\includegraphics[scale=0.1,angle=0]{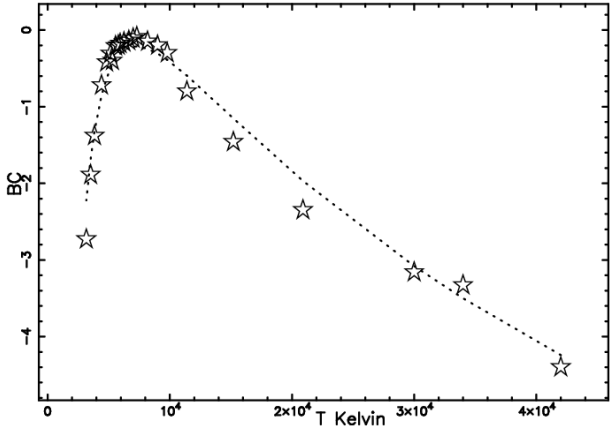}
}
\caption{Relationship  of BC against T, MAIN SEQUENCE V.
         Error on BC=0.025.
         The stars represent the calibrated data extracted
         from     Table 15.7 in~\cite{Cox} and
         the dotted line, the suggested fit.}
\label{f02}
\end{figure}

\begin{table}[h]
\caption{Table of coefficients from calibrated data.}
\label{coefficients}
\begin{tabular}{lccc}
\hline
  & MAIN, V  & GIANTS, III, & SUPERGIANTS I       \\
                   \hline
$K_{\mathrm{BV}}$       &-0.641 $\pm$ 0.01 &  -0.792 $\pm$ 0.06   & -0.749 $\pm$ 0.01 \\
$T_{\mathrm{BV}}[\mathrm {K}]$       &7360   $\pm$  66  &     8527$\pm$ 257    & 8261   $\pm$ 67   \\
$K_{\mathrm{BC}}$       & 42.74 $\pm$ 0.01 &  44.11  $\pm$ 0.06   & 42.87  $\pm$ 0.01 \\
$T_{\mathrm {BC}}[\mathrm {K}]$       & 31556 $\pm$  66  & 36856   $\pm$ 257    & 31573  $\pm$ 67   \\
$a_{\mathrm {LM}}$       &0.062 $\pm$ 0.04 &  0.32   $\pm$ 0.14    & 1.29   $\pm$ 0.32 \\
$b_{\mathrm {LM}}$       &3.43  $\pm$ 0.06 &  2.79   $\pm$ 0.23    & 2.43   $\pm$ 0.26 \\
\hline
\end{tabular}
\end{table}
These numerical results
can be visualised   in
Figure~\ref{f01}
when  (B-V) against  T is considered
and in Figure~\ref{f02}
where the  relationship between BC and T is reported;
see~\cite{Press} about the fitting techniques.

Other
authors analyse  the
form T=T((B-V), [Fe/H], $\log_{10}$~g)
where [Fe/H] represents the
metallicity and g the surface gravity, see~\cite{Sekiguchi}, or
a two piecewise function for  B-V =f (T), see for
example~\cite{Reed}.

The luminosity of the star is represented through the usual formula
\begin{equation}
\log_{10}(\frac {L}{L_{\sun}})  =  0.4 (4.74 -M_{\mathrm{bol}})
\quad,
\end {equation}
see,  for example, equation~(5.120) in~\cite{Lang}.
The connection with the mass is made through the numerical
relationship
\begin{eqnarray}
\log_{10}(\frac {L}{L_{\sun}})  = a_{\mathrm {LM}} +b_{\mathrm {LM}}\log_{10}(\frac {{\mathcal M}}
 {{\mathcal M}_{\sun}}) \\
 for~{\mathcal M} > 0.2 {\mathcal {M}}_{\sun} \nonumber
\quad.
\end {eqnarray}
The numerical coefficients $a_{\mathrm {LM}}$ and  $b_{\mathrm {LM}}$,
as  deduced from the calibrated data in Table 3.1 of~\cite{Deeming}
are reported in Table~\ref{coefficients}.

With the theory here adopted  the mass of the star is
\begin{eqnarray}
  \log_{10} \frac {{\mathcal M}}
 {{\mathcal M}_{\sun}}=   \nonumber
 \end{eqnarray}
 \begin{eqnarray}
\frac {- 0.4\,{\it M_{\mathrm V}}- 0.4\,{\it K_{\mathrm {BC}}}+ 4.0\,
 \ln
 \left( {\frac {{\it T_{\mathrm{BV}}}}{{\it (B-V)}-{\it K_{\mathrm{BV}}}}} \right)  \left( \ln
 \left( 10 \right)   ^{-1}
  \right )}{b_{\mathrm {LM}}}  \nonumber
  \end{eqnarray}
 \begin{eqnarray}
 -\frac{
  \ 0.4\,{\frac {{\it T_{\mathrm {BC}}}\,
 \left( {\it -(B-V)}+{\it K_{\mathrm{BV}}} \right)}{{\it T_{\mathrm{BV}}}}}+ 
1.896-{\it a_{\mathrm {LM}}
}} {b_{\mathrm {LM}}}
\label{mass_analytical}
\quad.
\end {eqnarray}

From a practical point  of view  the percentage  of
reliability  of our evaluation   can also be  introduced:
\begin{equation}
\epsilon  =(1- \frac{\vert( \mathcal {M}^{\mathrm {cal}}- 
{\mathcal M}^{\mathrm {num}}) \vert}
{\mathcal{M}^{\mathrm {cal}}+\mathcal{M}^{\mathrm {num}}}) \times 100
\,,
\label{efficiency}
\end{equation}
where $\mathcal {M}^{\mathrm {cal}}$ is the   calibrated value of the star
mass, see Table 15.8 in~\cite{Cox},
and $\mathcal {M}^{\mathrm {num}}$ is
the numerical value here evaluated.
The resulting masses are reported in
Table~\ref{table_masses_analytical}.

\begin{table}[h]
\caption{Table of evaluated masses through an analytical method,
  equation~(\ref{mass_analytical}).}
\label{table_masses_analytical}
\begin{tabular}{lcccc}
\hline
sp  & MK Class &$\mathcal {M}^{\mathrm {cal}}/\mathcal{M}_{\sun}$ & $\mathcal
{M}^{\mathrm {num}}/\mathcal{M}_{\sun}$ &
$ \epsilon (\%)$      \\
                   \hline
B0  & V    & 17.5 & 17.34   & 99.54       \\
A0  & V    & 2.9  & 3.42    & 91.7        \\
F0  & V    & 1.6  & 1.76    & 95.23       \\
G0  & V    & 1.05 & 1.13    & 96.02      \\
K0  & V    & 0.79 & 0.81    & 98.73      \\
M2  & V    & 0.4  & 0.38    & 98.38       \\
K5  & III  & 1.2  & 6.38    & 31.63       \\
M0  & III  & 1.2  & 7.15    & 28.71      \\
A0  &  I   & 16.0 & 22.35   & 83.44        \\
M0  &  I   & 13.0 & 27.73   & 63.82       \\
M2  &  I   & 19.0 & 28.6    & 79.82       \\
\hline
\end{tabular}
\end{table}
The formulae here derived can be used
to deduce the mass  of the stars once the parameters
$M_{\mathrm V}$, B-V and luminosity class  are provided.

\section{Masses}

We now apply the developed techniques to the nearest stars
and to an open cluster, NGC6649.

\label{sec_masses}
\subsection{The masses in the first 10~pc}

The completeness in masses of the sample can be evaluated 
in the following way.
The limiting apparent magnitude 
$m_v^L$  is known, for example  in the case of Hipparcos it
is  8. 
The corresponding absolute limiting magnitude is computed 
and inserted in equation~(\ref{mass_analytical}).  The limiting 
mass , ${\mathcal M^L}$  , 
for stars belonging to the MAIN SEQUENCE V  is  
\begin{eqnarray}
\log_{10} \frac {{\mathcal M^L}}
 {{\mathcal M}_{\sun}}=   \nonumber
 \end{eqnarray}
 \begin{eqnarray}
- 0.1164\,{\it m^l_{\mathrm V}}+ 0.2528\,\ln  \left(  0.1\,{\it d_1} \right) -
 4.122+
\nonumber 
\end{eqnarray}
\begin{eqnarray}
 0.505\,\ln  
\left( 
\frac {0.368\, 10^{11}} 
      {0.5\, 10^7\,(B-V)+ 0.3206\,10^7}  
\right) +
 0.499\,{\it (B-V)}
\label{mass_limit}
\quad ,
\end {eqnarray}
where $d_1$   is the maximum distance that 
characterises the sample in pc.
When 
$d_1$=10 , 
$m_v^L$=8 ,
and (B-V)=1.15 
we obtain 
$\mathcal {M}^L = 0.53\mathcal {M}_{\sun}$;
this is the limit over which the sample in masses  is complete.

\begin{figure}
\resizebox{\hsize}{!}
{
\includegraphics[scale=0.1,angle=0]{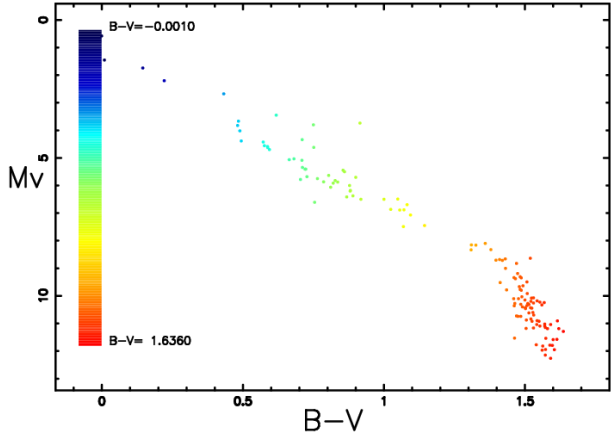}
}
\caption{$M_{\mathrm V}$  against (B-V)   (H-R diagram) in the first  10~pc,
    this sample is incomplete in masses.
         }
\label{f03}
\end{figure}

Due to the fact that Hipparcos (\cite{Esa})
gives B-V, $m_{\mathrm V}$ and parallax,
we can implement our algorithm on that  database.
The classification in MK classes (I,III,V) is then obtained
by a comparison with the interval of existence of 
 calibrated
stars.
In Figure~\ref{f03} we report the H--R diagram
of the first 10~pc when the incomplete (in masses)
 sample is considered.

\begin{figure}
\resizebox{\hsize}{!}
{
\includegraphics[scale=0.1,angle=0]{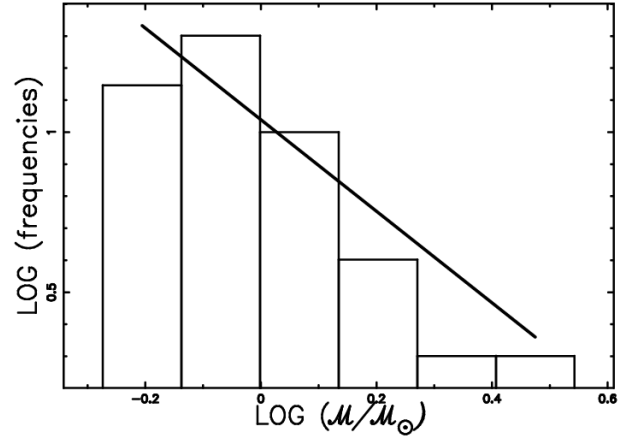}
}
\caption{
Log--Log histogram  of  mass distribution
with a superposition of the
fitting line (the power law  distribution)
when the complete sample is considered.
Parameters as in Table~\ref{stars}.
}
\label{f04}
\end{figure}

\begin{figure}
\resizebox{\hsize}{!}
{
\includegraphics[scale=0.1,angle=0]{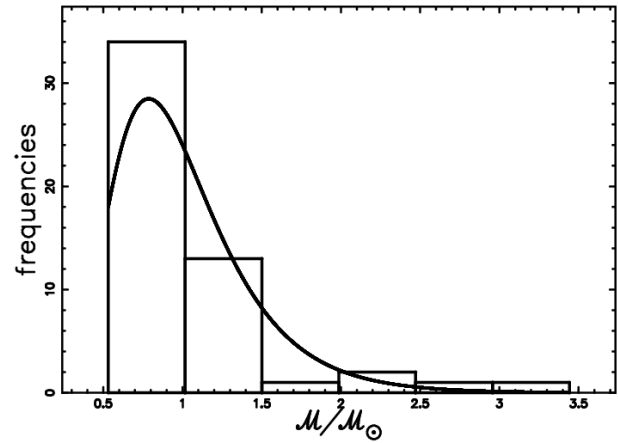}
}
\caption{
Histogram (step-diagram)  of  mass distribution
with a superposition of the
lognormal  distribution
when the complete sample is considered.
Parameters as in Table~\ref{stars}.
}
\label{f05}
\end{figure}

In Table~\ref{stars} we report
the coefficients of the three  pdf,
average value,
the number, $N_s$,  of stars and
the  merit function $\chi^2$.
A typical fit of the complete sample
in the light of the  power law   pdf is reported in
Figure~\ref{f04}, conversely
Figure~\ref{f05} reports the fit through
the lognormal distribution.
\begin{table}[h]
\caption{Coefficients of  mass distribution
         in the first 10~pc       , 
         of  the incomplete sample,
         GIANTS included          ,
          $\mathcal {M} \geq  0.2\mathcal {M}_{\sun}$,
          $\overline {\mathcal {M}}= 0.61 \mathcal {M}_{\sun}$
          and $N_s$=137.}
\label{stars}
\begin{tabular}{lccc}
\hline
\hline
Kiang      &  $\chi^2$ =14.28  &   c=    1.67     &        \\
\hline
Power law  &  $\chi^2$ =17.6   & $\alpha$=1.85    &         \\
\hline
lognormal  &  $\chi^2$ =0.95   & $\mu$ =0.6       &$\sigma$= 0.57  \\
\hline
\hline
\end{tabular}
\end{table}

\begin{table}[h]
\caption{Coefficients of  mass distribution
         in the first 10~pc,  of
         the complete sample,
          $\mathcal {M} \geq  0.53\mathcal {M}_{\sun}$,
          $\overline {\mathcal {M}}= 0.98  \mathcal {M}_{\sun}$,
          GIANTS excluded , 
          and $N_s$=57.}
\label{stars_complete}
\begin{tabular}{lccc}
\hline
\hline
Kiang      &  $\chi^2$ =12.62  &   c= 3.42        &        \\
\hline
Power law  &  $\chi^2$ =8.7   & $\alpha$=2.42    &         \\
\hline
lognormal  &  $\chi^2$ =4.55   & $\mu$ =1.01       &$\sigma$= 0.41 \\
\hline
\hline
\end{tabular}
\end{table}

\subsection{The masses in open clusters}

The distribution of  masses in open clusters can be considered
a further application of the pdf here studied.
We briefly review some of the typical clusters subject
to  investigation:
IC348
( \cite{Preibisch}  ;
\cite{Muench}       ;
\cite{Luhman}       ),
Taurus
(  \cite{Luhman_2000} ;
   \cite{Briceno}    ),
Orion trapezium
( \cite{Hillenbrand}  ;
  \cite{Luhman_b}     ;
  \cite{Muench_b}     ),
Pleiades
( \cite{Bouvier} ;
  \cite{Luhman_b}), and
M35
(\cite{Barrado} ).

\begin{figure}
\resizebox{\hsize}{!}
{
\includegraphics[scale=0.1,angle=0]{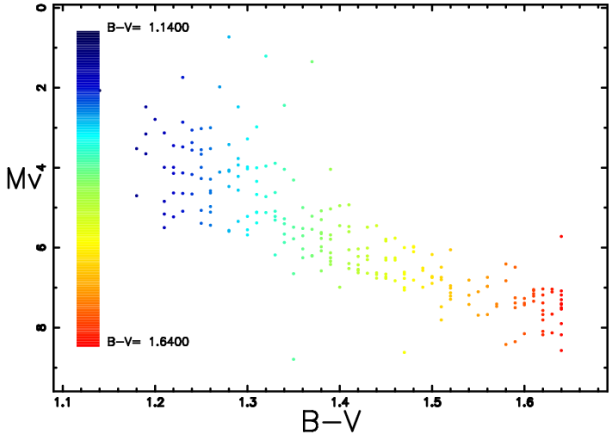}
}
\caption{$M_{\mathrm V}$  against (B-V)   (H-R diagram) in NGC6649.
         }
\label{f06}
\end{figure}

Here attention is focused on NGC6649, see~\cite{Walker}.
In this case the distance  modulus is 11 (1585~pc) and this
fact allows the determination of the  masses
from $M_{\mathrm V}$--(B-V) diagram, see Figure~\ref{f06},
once the
algorithm outlined in Sect.~\ref{masses} is adopted.

\begin{figure}
\resizebox{\hsize}{!}
{
\includegraphics[scale=0.1,angle=0]{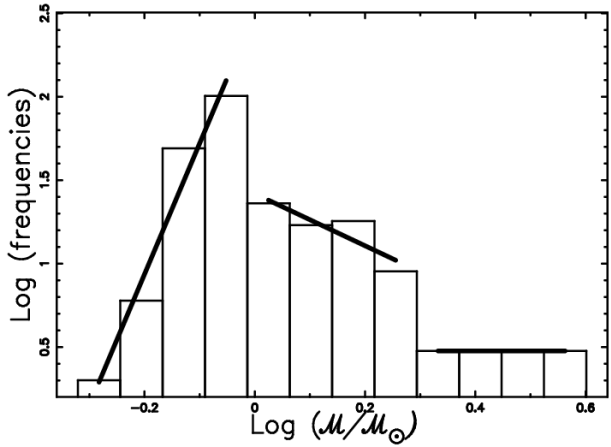}
}
\caption{
Log--Log histogram  of  mass distribution in NGC6649
with a superposition of the three--part power law.
Parameters as in Table~\ref{trezone}.
}
\label{f07}
\end{figure}

In Table~\ref{NGC6649} we report the parameters relative to the
three pdf.
Once the framework of the
three power law is accepted~,
Table~\ref{trezone}  reports the results
that are visualised in Figure~\ref{f07}.
\begin{table}[h]
\caption{Coefficients of mass distribution
         in NGC~6649 where
            $N_s$=234 and
            $\overline {\mathcal {M}}= 1.04  \mathcal {M}_{\sun}$.}
\label{NGC6649}
\begin{tabular}{lccc}
\hline
\hline
Kiang      &  $\chi^2$ =35.44  &   c=5.02         &        \\
\hline
Power law  &  $\chi^2$ = 248.49  & $\alpha$=3.12    &         \\
\hline
lognormal  &  $\chi^2$ =29.39   & $\mu$ =1.03       &$\sigma$=0.32  \\
\hline
\hline
\end{tabular}
\end{table}
%
%
\begin{table}[h]
\caption{Three power law
         in NGC~6649 where
          $N_s$=234.}
\label{trezone}
\begin{tabular}{lcc}
\hline
\hline
Interval        & $\alpha$    \\
\hline
0.52 $\leq{\mathcal M}/{\mathcal M}_{\sun}  <$ 0.88   & - 6.83  \\
0.88 $\leq{\mathcal M}/{\mathcal M}_{\sun}  <$ 1.8  &  2.55   \\
1.80 $\leq{\mathcal M}/{\mathcal M}_{\sun}  <$3.65 & 1.0     \\
\hline
\hline
\end{tabular}
\end{table}

\begin{figure}
\resizebox{\hsize}{!}
{
\includegraphics[scale=0.1,angle=0]{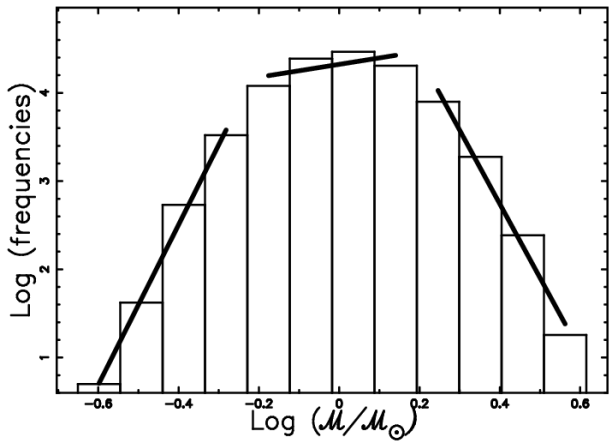}
}
\caption{
Log--Log histogram  of  100000 masses
randomly generated according to  lognormal distribution
(parameters as in Table~\ref{trezone})
with a superposition of the three--part power law.
}
\label{f08}
\end{figure}

\section {Conclusions}

A new method  represented by 
equation~(\ref{mass_analytical}), allows us to deduce the mass
of  stars from photometric data.
This method is limited in mass , see equation~(\ref{mass_limit}) , 
in the same way as the 
Hipparcos  data are  complete up to $m_v$=8.

In the second part of this paper the obtained  mass distribution
is fitted with  the pdf of  physical nature: 
the lognormal and the gamma
variate.
The gamma-variate  is applied for the first time as 
a fitting function of  mass distribution.
When the complete sample of masses is fitted   with
the gamma-variate , c=3.42 , 
is found ; a comparison should be made with c=4 , which  
is the value suggested by~\cite{Kiang} in the case of 2D fragmentation.
Careful analysis of the results of Table~\ref{NGC6649}  and
Table~\ref{stars_complete}  indicates that the lognormal 
gives the best fit to the star's masses. 
On the contrary the three power law
function  simply represents the various segments of a
continuous curve. 
This point of view can be expressed  by plotting
100000  stars  generated according to  lognormal data, 
see
Figure~\ref{f08}.

In Figure~\ref{f08}~,
 the three indices
$\alpha_i$ are now varying in a continuous way.
The  great variations in $\alpha_i$ visible in Table~\ref{trezone}
and Figure~\ref{f07},
 can be due to the transition
from a great number of stars ( smooth behaviour of $\alpha_i$ )
to a small number of stars   ( jumps in $\alpha_i$ or
variability in the IMF ).

\begin{figure}
\resizebox{\hsize}{!}
{
\includegraphics[scale=0.1,angle=0]{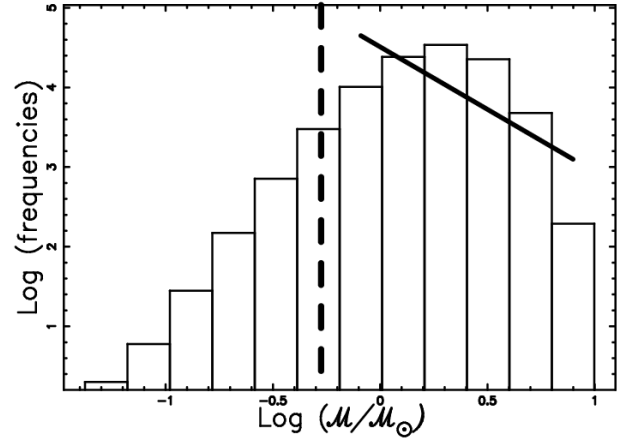}
}
\caption
{
Log--Log histogram  of  100000 masses
randomly generated according to the Kiang distribution
with c=4  and  
$\overline {\mathcal {M}}= 2 \mathcal {M}_{\sun}$.
On the right , $\mathcal {M} \geq  0.51 \mathcal {M}_{\sun}$, 
is visible the fitting  power law distribution  
with $\alpha$=2.56.
The vertical dashed line represents the limiting mass 
$\mathcal {M}^L = 0.53\mathcal {M}_{\sun}$.
}
\label{f09}
\end{figure}

It is now possible in the light of the limiting mass $\mathcal {M}^L$, see 
equation~(\ref{mass_limit}) , to explain the reason for which 
a power law fits  the masses well  in the first 10 pc.
The limiting mass introduces a cut in 
the physical sample 
at a value that is  probably near the averaged value,
 and therefore
only the masses of  value greater than the averaged value are detected.
This point of view is simulated in Figure~\ref{f09} in which
a real continuous mass distribution as given by a gamma-variate with
c=4 is reported:  due to the limiting mass $\mathcal {M}^L$ a power law 
distribution is observed.

\section {acknowledgements}
{This note amplifies  a series of 
lessons given  in  the
course "Fondamenti di Astrofisica" 
on behalf of
the organiser 
Roberto Gallino
at the Turin University.
I am grateful to the referee of this paper for 
helpful suggestions. 
}

\end{document}